\documentclass[prb,twocolumn,showpacs,superscriptaddress]{revtex4}

\usepackage{graphicx}

\begin{document}

\title{Effect of spatial inhomogeneity on the mapping between strongly 
interacting fermions and weakly interacting spins}

\author{Vivian V. Fran\c{c}a}\email{vivian.franca@physik.uni-freiburg.de}
\affiliation{Physikalisches Institut, Albert-Ludwigs Universit\"at, Hermann-Herder-Str.3, Freiburg, Germany}
\affiliation{Instituto de F\'{\i}sica de S\~ao Carlos,
Universidade de S\~ao Paulo, S\~ao Carlos, 13560-970 S\~ao Paulo,
Brazil}

\author{Klaus Capelle}
\affiliation{Centro de Ci\^encias Naturais e Humanas,
Universidade Federal do ABC, Santo Andr\'e, 09210-170 S\~ao Paulo,
Brazil}
\affiliation{Instituto de F\'{\i}sica de S\~ao Carlos,
Universidade de S\~ao Paulo, S\~ao Carlos, 13560-970 S\~ao Paulo,
Brazil}

\date{\today }

\begin{abstract}
A combined analytical and numerical study is performed of the mapping
between strongly interacting fermions and weakly interacting spins, in the
framework of the Hubbard, t-J and Heisenberg models. While for spatially 
homogeneous models in the thermodynamic limit the mapping is thoroughly 
understood, we here focus on aspects that become relevant in spatially
inhomogeneous situations, such as the effect of boundaries, impurities,
superlattices and
interfaces. We consider parameter regimes that are relevant for traditional 
applications of these models, such as electrons in cuprates and manganites, 
and for more recent applications to atoms in optical lattices. The rate of 
the mapping as a function of the interaction strength is determined from the
Bethe-Ansatz for infinite systems and from numerical diagonalization for 
finite systems. We show analytically that if translational symmetry is broken 
through the presence of impurities, the mapping persists and is, in a certain 
sense, as local as possible, provided the spin-spin interaction between two 
sites of the Heisenberg model is calculated from the {\em harmonic mean} of 
the onsite Coulomb interaction on adjacent sites of the Hubbard model.
Numerical calculations corroborate these findings also in interfaces and superlattices,
where analytical calculations are more complicated.
\end{abstract}


\newcommand{\be}{\begin{equation}}
\newcommand{\ee}{\end{equation}}
\newcommand{\bea}{\begin{eqnarray}}
\newcommand{\eea}{\end{eqnarray}}
\newcommand{\bi}{\bibitem}
\newcommand{\la}{\langle}
\newcommand{\ra}{\rangle}
\newcommand{\ua}{\uparrow}
\newcommand{\da}{\downarrow}
\renewcommand{\r}{({\bf r})}
\newcommand{\rp}{({\bf r'})}
\newcommand{\eps}{\epsilon}
\newcommand{\bfr}{{\bf r}}

\maketitle

%\tableofcontents

\section{Introduction}
\label{intro}

Strongly interacting fermions are part of some of todays most studied physical 
systems. In cuprate and manganite systems, for example, strongly correlated 
electrons are held responsible for high-temperature superconductivity and 
colossal magnetoresistance, respectively.\cite{electrons1,electrons2,electrons3}
Systems of strongly interacting fermionic atoms can be realized in optical lattices, 
and are currently under intense investigation due to the possibility to use 
them as quantum simulators for understanding phenomena of 
condensed matter physics.\cite{atoms1,atoms2,atoms3}

Since the seminal works of Wigner on the low-density electron 
crystal\cite{wigner} and of Mott on the metal-insulator 
transition\cite{mott} it is known that strong repulsive particle-particle 
interactions suppress itineracy and favor localization.\cite{mit} 
In the localized state, the repulsive interaction is minimized, and charge 
degrees of freedom are frozen out. The dominating interactions in this state 
are magnetic.

Mathematically, strongly interacting fermions are frequently described by the
Hubbard model, which in one dimension is defined by the Hamiltonian
\begin{equation}
\hat{H}^{Hubb}=-t\sum_{i\sigma}(\hat{c}_{i\sigma}^\dagger\hat{c}_{i+1,\sigma}+H.c.)+ U \sum_i \hat{n}_{i\uparrow}\hat{n}_{i\downarrow},
\label{hub}
\end{equation}
where $\hat{c}_{i\sigma}^\dagger$ and  $\hat{c}_{i\sigma}$ are fermionic
creation and destruction operators, $\hat{n}_{i\sigma}=\hat{c}_{i\sigma}^\dagger\hat{c}_{i\sigma}$ is the particle-density operator, $U$ the onsite
interaction and $t$ the hopping parameter.

For sufficiently strong interactions, the Hubbard model can be expanded in 
powers of $t/U$. (Below we quantify what interactions can be considered 
`sufficiently strong'.) The leading term of this expansion is the t-J model,
\bea
\hat{H}^{tJ}&=&-t\sum_{i\sigma}(\hat{c}_{i\sigma}^\dagger\hat{c}_{i+1,\sigma}+H.c.)\nonumber\\
&&+\frac{4t^2}{U}\sum_{i}\left[\vec{\hat{S}}_i\cdot \vec{\hat{S}}_{i+1} -\frac{\hat{n}_i\hat{n}_{i+1}}{4}\right],
\label{tj}
\eea
where $\vec{\hat{S}}_{i}$ is the spin one-half vector operator at each site. 
This model is frequently taken to be the starting point in investigations of 
doped cuprates.

For a half-filled system, in which the number of fermions $N$ equals the 
number of lattice sites $L$, the average density $n=N/L$ is unity. Since there 
are no empty sites, hopping is suppressed, and the t-J model reduces to the
antiferromagnetic Heisenberg model
\begin{equation}
\hat{H}^{Heis}=J\sum_{i}\left[\vec{\hat{S}}_i\cdot \vec{\hat{S}}_{i+1} -\frac{1}{4}\right],
\label{heis}
\end{equation}
where $J=4t^2/U$ and charge fluctuations are completely frozen out.
The original system of strongly interacting 
itinerant fermions ($U/t \gg 1$) has thus been mapped on a system of localized 
spins with weak antiferromagnetic interactions ($0 \leq J/t \ll 1$). 

The mathematics and the physics of this mapping are very well understood and
discussed in the textbook literature.\cite{auerbach,fulde} The mapping
of strongly interacting itinerant fermions on weakly interacting localized
spins is a standard concept of condensed-matter physics, routinely used in 
the interpretation of experiments on strongly correlated solids. 
Recently, however, three important classes of systems have been discovered 
or created that call for a reconsideration and more detailed investigation 
of this mapping.

It is well known that many strongly
correlated systems are characterized by nanoscale spatial inhomogeneity. 
Such inhomogeneity can take the form of irregular spatial variations of
system properties, such as observed by scanning-tunneling microscope
techniques in many cuprates and similar materials,\cite{electrons2,inhom0,inhom1,inhom2,inhom3,inhom4} or the form of regular spatial variations 
such as in naturally occurring or man-made superlattice structures.\cite{paiva,malvezzi,malvezzi2,superl3,superl4} In the presence of either type of inhomogeneity, the 
parameters characterizing the model Hamiltonian become site dependent. In
the simplest case, with which we are mostly concerned here, the above
homogeneous Hubbard model is replaced by an inhomogeneous model of the form
\begin{equation}
\hat{H}^{Hubb}_{inh}=-t\sum_{i\sigma}(\hat{c}_{i\sigma}^\dagger\hat{c}_{i+1,\sigma}+H.c.)+ \sum_i U_i \hat{n}_{i\uparrow}\hat{n}_{i\downarrow},
\label{inhomhub}
\end{equation}
in which the onsite interaction $U_i$ varies from site to site. 

A second class of systems we are concerned with here are nanoscale devices.
In the modeling of such devices inhomogeneities in the model parameters occur 
simply because on the nanoscale the effect of the surface can no longer be 
neglected, and also because a typical device combines more than one material, 
with the resulting interface automatically implying the existence of spatial 
variations in the system parameters.

Finally, in still another line of research, ultracold atom gases have been 
trapped optically and arranged in optical lattices.\cite{atoms1,atoms2} 
In optical traps the system parameters can be controlled and varied in ways 
not possible in solid-state situations. In particular, the onsite interaction
$U_i$ can attain values $U/t \approx 100$ or larger. Such values are way 
beyond what is considered `strongly correlated' in solid-state physics.

Motivated by all these systems, we present, in the present paper, a combined
analytical and numerical study of the mapping from the Hubbard model to
the Heisenberg model in the presence of spatial inhomogeneity.
In Sec.~\ref{rate} we investigate the rate of the mapping, as measured
by the difference in ground-state energies of the Hamiltonians. We allow 
the interaction strength to go beyond its typical solid-state values and to 
enter the ultrastrong regime attainable for cold atoms. 

In Sec.~\ref{inh} we turn to our main subject, the inhomogeneous Hubbard 
model of Eq.~(\ref{inhomhub}). 
In Sec.~\ref{analyt1} we investigate analytically the case of a single impurity, described 
as one site with a value of $U$ differing 
from all others, and show that the Hubbard-to-Heisenberg mapping is preserved essentially
in its homogeneous form, provided the effective $J$ is calculated from 
the {\em harmonic mean} of the values of $U$ on the sites connected by $J$. 
We illustrate this finding numerically, by contrasting,
for an impurity system, results obtained from the harmonic mean with results
obtained from the arithmetic, geometric and quadratic means. 
In Sec.~\ref{complexinhom} we 
show that the harmonic mean allows to extend the Hubbard-to-Heisenberg mapping 
to systems with more complicated types of spatial inhomogeneity, such as 
superlattices, disordered systems and interfaces between different materials. 
Section~\ref{concl} contains our conclusions.

\section{Rate of the mapping for translationally invariant  systems}
\label{rate}

In a first step, to provide the background for the later investigations, we
consider spatially homogeneous infinite Hubbard and Heisenberg chains, and
investigate the rate of the approach of the ground-state energies of both models as a function of
$U$. This allows us to quantify the rate at which charge fluctuations are frozen out.

\begin{figure}[t!]
\includegraphics[height=6cm]{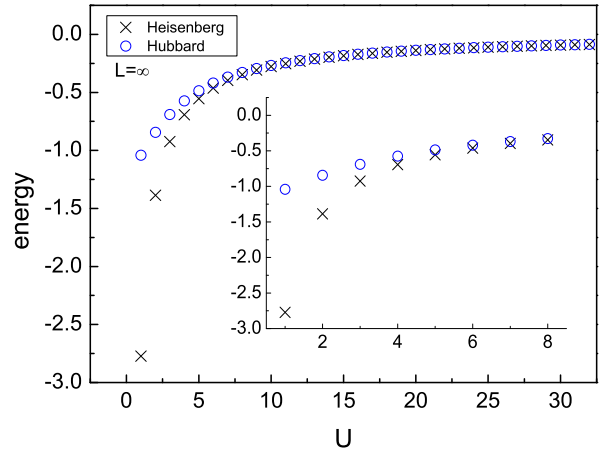}
\caption{Per-site GS energy as a function of interaction in the thermodynamic 
limit. Comparison between the Heisenberg with $J=4t^2/U$ and
the half-filled Hubbard chain. The main figure includes values of $U$ that 
can be realized in systems of trapped cold atoms, while the inset displays 
results for values of $U$ typical of weakly and strongly correlated solids.}
\label{fig1}
\end{figure}

The per-site ground-state (GS) energy of the Heisenberg chain in the 
thermodynamic limit is  
\be
e^{Heis}(J)= \lim_{L\to \infty}{E^{Heis} \over L}=-J\ln(2).
\ee
The per-site GS energy of infinite half-filled ($n=1$) Hubbard chain at 
$U/t \to \infty$ is  
\bea
e^{Hubb}(n=1,{U\over t} \to\infty)
= \lim_{L\to \infty}{E^{Hubb} \over L} 
= - {4t^2\over U} \ln(2).
\eea
Both expressions become identical for $J = 4t^2/U$. In
order to quantitatively investigate the mapping at finite $U/t$, we 
calculate $e^{Hubb}(n=1,U)$ by numerically solving the Bethe-Ansatz integral
equations\cite{liebwu,shiba} as a function of $U$ and compare the result to 
the energy of the Heisenberg model at the corresponding value of $J$, 
{\em i.e.} $e^{Heis}(J=4t^2/U)$. The result is displayed\cite{foot1} in Fig.~\ref{fig1}. 

In Table~\ref{tab1} we show the relative percentage deviation between the 
GS energies, 
\be
D(\%)=100\frac{e^{Hubb}-e^{Heis}}{e^{Hubb}},
\ee
for various representative values of $U$. 
This comparison between both models becomes trivial, once the Bethe-Ansatz solution is available,
but already leads to a first somewhat unexpected conclusion. Frequently,
the Heisenberg model is taken to be the starting point for a description 
of undoped antiferromagnetic insulating parent compounds of high-temperature
superconductors. The effect of doping is accounted for by going from the
Heisenberg to the t-J model, arguing that the latter should be a reasonable
approximation to the Hubbard model for the involved large values of $U$.

What the comparison in Fig.~\ref{fig1} and Table~\ref{tab1} shows is that 
for values of $U$ that are representative of cuprate materials the
t-J or Heisenberg models provide at best a semiquantitative approximation
to the Hubbard model. At $U=6$ the difference between both ground-state 
energies is approximately $10\%$. Charge fluctuations are thus not yet frozen
out for such $U$, even at half filling. The rather large deviation observed shows 
that the mapping of strongly interacting fermions onto weakly
interacting spins is not quantitatively reliable for, e.g., cuprate systems at 
realistic values of $U$. There 
is no doubt, of course, that the t-J model captures the correct physics of the 
large-U Hubbard model -- the above questioning only refers to the accuracy
to which one needs to obtain a solution of the former, given that in the
parameter regime typical of strongly-correlated solids it is
itself only a moderately accurate representation of the latter.

\begin{table}[t!]
\caption{Relative percentage deviation of the per-site GS energies for values 
of $U$ typical of weakly correlated solids ($U=1$), strongly correlated solids
($U=6$) and values that are attainable for atoms in optical traps.}
\begin{ruledtabular}
\begin{tabular}{c|ccccccc}
$U$ & 1 & 6 & 10 & 20 & 50 & 100 & 200 \\
\hline
$D(\%)$ &-166.50 & -10.01 & -3.78 & -0.97 & -0.16 & -0.04& -0.01\\
\end{tabular}
\end{ruledtabular}
\label{tab1}
\end{table}

We note that this analysis is based on GS energies.
An alternative comparison between both Hamiltonians would proceed in terms of
the overlap of their wave functions, instead of the difference of their
energies. Our main interest in this initial investigation is to investigate
for which values of $U$ the mapping breaks down, and for this it is
enough to find one quantity that is not properly reproduced. Thus, if we use
our analysis to indicate when the mapping does {\em not} hold, we
are on the safe side by using energies. Still another, and indeed more
fundamental, mode of analysis proceeds by directly comparing the Hamiltonians.
We use this procedure in Sec.~\ref{analyt1}, where our interest is not only in
when the mapping breaks down, but also in how it can be restored.

\section{Mapping in the presence of spatial inhomogeneity}
\label{inh}

We now turn to systems where the inhomogeneity occurs not only at the surface, due to 
finite size, but also in the bulk. A typical case is that of
a localized impurity or defect, modeled by one site with onsite interaction
differing from that of all the others. We take this simple system as 
representative of Hubbard models with broken translational symmetry, and
focus most of our analysis on it. The extension of our conclusions to
interfaces and superlattices is discussed in Sec.~\ref{complexinhom}.

Intuitively, one would expect that a localized perturbation of the 
homogeneous Hubbard model should only produce a similarly localized
perturbation of the homogeneous Heisenberg model. However, the relation
between both models involves a projection on the subspace with no double 
occupation, together with an expansion in inverse 
powers of $U$,\cite{auerbach,fulde} and it is not clear from the outset to
what extent these operations preserve the above naive expectation of locality.

\begin{figure}[t!]
\centering
\includegraphics[viewport= 40 40 200 1000, angle=-90, width=13.2cm]{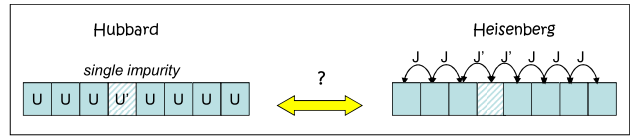}
\caption{Illustration of the inhomogeneous Hubbard model with a single
impurity $U' \neq U$, and its putative relation to an inhomogeneous Heisenberg 
model with two bond defects $J' \neq J$.}
\label{fig2}
\end{figure}

In fact, there is one sense in which the mapping, if it continues to exist, 
cannot be local: $U$ is defined on one site, while $J$ connects two adjacent
sites. While this difference is almost irrelevant in the homogeneous case
where all sites are equivalent and translational symmetry rules, it becomes
important in the inhomogeneous case, where any change in the onsite $U$ 
on the Hubbard model must affect the corresponding intersite $J$ for at least 
two sites of the Heisenberg model. This is illustrated in Fig.~\ref{fig2}.

The questions to address are thus (i) whether the mapping still exists in 
the absence of translational symmetry, (ii) how to calculate the Heisenberg 
$J$ from the Hubbard $U$ in inhomogeneous systems, and (iii) if the mapping
is as local as possible, {\em i.e.}, involves only sites adjacent to the 
impurity site, or requires a higher degree of nonlocality.
In next subsections we address these questions 
analytical and numerically.

\subsection{A single impurity}
\label{analyt1}

We start with the one-dimensional Hubbard model in the presence of a single
impurity at site $k$ with onsite interaction $U'$ differing from the 
background value $U$ on all other sites $i\neq k$,  
\bea
\hat{H}^{Hubb}_{inh}&=&-t\sum_{i\sigma}(\hat{c}_{i\sigma}^\dagger\hat{c}_{i+1,\sigma}+H.c.)
+ U \sum_{i\neq k} \hat{n}_{i\uparrow}\hat{n}_{i\downarrow}
\nonumber \\
&&+ U'\hat{n}_{k\uparrow}\hat{n}_{k\downarrow}.\label{hubbard_inh}
\eea
The standard proof of the mapping from the Hubbard to the t-J model\cite{auerbach,fulde} can be repeated for this Hamiltonian and leads to the inhomogeneous
t-J model with Hamiltonian 
\begin{eqnarray}
\hat{H}_{inh}^{tJ}&=&-t\sum_{i\sigma}(\hat{c}_{i\sigma}^\dagger\hat{c}_{i+1,\sigma}+H.c.)\hspace{1cm}\\
&&+\frac{4t^2}{U}\sum_{i\neq k}\left[\vec{\hat{S}}_i\cdot \vec{\hat{S}}_{i+1} -\frac{\hat{n}_i\hat{n}_{i+1}}{4}\right]\nonumber\\
&&+\frac{4t^2}{U'}\left[\vec{\hat{S}}_k\cdot \vec{\hat{S}}_{k+1} -\frac{\hat{n}_k\hat{n}_{k+1}}{4}\right],\nonumber
\end{eqnarray}
where we assumed that both $U$ and $U'$ are much larger than $t$. We rewrite
this Hamiltonian by extracting one term from the sum over $i\neq k$
to obtain
\bea
\hat{H}_{inh}^{tJ}&=&-t\sum_{i\sigma}(\hat{c}_{i\sigma}^\dagger\hat{c}_{i+1,\sigma}+H.c.) \\
&&+\frac{4t^2}{U}\sum_{i\neq k,l}\left[\vec{\hat{S}}_i\cdot \vec{\hat{S}}_{i+1} -\frac{\hat{n}_i\hat{n}_{i+1}}{4}\right]\nonumber\\
&&+\frac{4t^2}{U}\left[\vec{\hat{S}}_k\cdot \vec{\hat{S}}_{k+1} -\frac{\hat{n}_k\hat{n}_{k+1}}{4}\right]
\nonumber \\
&&+\frac{4t^2}{U'}\left[\vec{\hat{S}}_{l}\cdot \vec{\hat{S}}_{l+1} -\frac{\hat{n}_{l}\hat{n}_{l+1}}{4}\right].\nonumber
\label{inhomtj}
\eea 
Now we take the average density $n=N/L=1$. {\it A priori} this average can be obtained from many different distributions $n_i$. However, since we are already in the limit $U,U'\gg t$, the total interaction energy on the background and the impurity sites is minimized by the particular distribution $n_i=1$ for all $i$. Deviations from this are due to hopping processes that become increasingly suppressed as $U$ and $U'$ grow. Thus the Hamiltonian (\ref{inhomtj}) at $n=1$ reduces to
\begin{eqnarray}
\hat{H}_{inh}^{tJ}(n=1)&=&+\frac{4t^2}{U}\sum_{i\neq k,l}\left[\vec{\hat{S}}_i\cdot \vec{\hat{S}}_{i+1} -\frac{1}{4}\right]\label{tj.final}\\
 &&+4t^2\left(\frac{1}{U}+\frac{1}{U'}\right)\left[\vec{\hat{S}}_l\cdot \vec{\hat{S}}_{l+1} -\frac{1}{4}\right]\nonumber,
\end{eqnarray}
where we chose $l$ as a neighbor site of $k$, $k=l+1$, such that $\vec{\hat{S}}_l\cdot \vec{\hat{S}}_{l+1}=\vec{\hat{S}}_k\cdot \vec{\hat{S}}_{k+1}$. This, in turn, can be written as
\begin{eqnarray}
H^{Heis}_{inh}=
J\sum_{i\neq l,l+1}\left[\vec{\hat{S}}_i\cdot \vec{\hat{S}}_{i+1} -\frac{1}{4}\right]+ 2J'\left[\vec{\hat{S}}_{l}\cdot \vec{\hat{S}}_{l+1} -\frac{1}{4}\right]\nonumber,\\
\label{inhomhe}
\end{eqnarray}
which has the form of a spatially inhomogeneous Heisenberg model with
background interaction $J=4t^2/U$ and two bond defects $J'=4t^2/\bar{U}^H$, 
where $\bar{U}^H$ is the harmonic mean of $U$ and $U'$,
\be
\bar{U}^H = \frac{2UU'}{U+U'}.
\label{harm}
\ee

This derivation tells us that the Hubbard model with a single impurity can
indeed be mapped onto a Heisenberg model with two bond defects, provided the
impurity site and the background sites both have repulsive interactions 
that are much larger than $t$, and that the $J$ connecting the impurity site
with its neighbors to the left and to the right is calculated from the
{\em harmonic mean} of the two onsite interactions. The mapping then still 
exists and is seen to be {\em as local as possible}, in the sense explained 
above.

In order to investigate the rate of the mapping in inhomogeneous systems, we now perform a numerical investigation of both models. For illustration we also include in these calculations the quadratic, arithmetic and geometric means,
\begin{eqnarray}
\bar{U}^Q&=&\sqrt{\frac{U^2+U'^2}{2}}\label{q}\\
\bar{U}^A&=&\frac{U+U'}{2}\label{a}\\
\bar{U}^G&=&\sqrt{UU'}\label{g}.
\end{eqnarray}
Both the inhomogeneous Hubbard model with $n=1$ and one $U'\neq U$ and the 
inhomogeneous Heisenberg model with two $J'\neq J$ are diagonalized numerically, 
and compared by means of the deviation between their GS energies.
We use the same method of analysis as in Sec.~\ref{rate}, this time, however,
applied to impurity systems.  

\begin{figure}[t!]
\includegraphics[height=10cm, width=7.5cm]{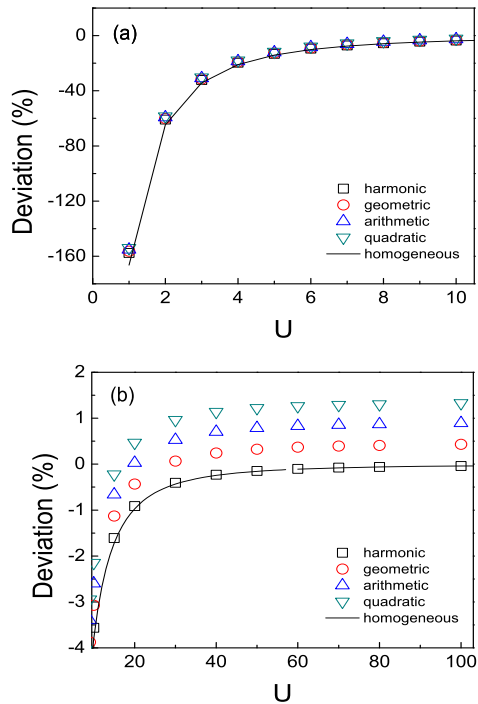}
\caption{Relative percentage deviation between the GS energies of the
single-impurity Hubbard chain and the two-defect Heisenberg chain, obtained
from numerical diagonalization, adopting $J'=J(\bar{U})$ with four
different choices for the average $\bar{U}$: (a) values of $U$ that are
typical of solids and (b) values of $U$ that are attainable in systems
of optically trapped atoms. System parameters: $L=8$ sites, $N=8$ fermions, open boundary
conditions, impurity strength $U'=3U/2$.}
\label{fig3}
\end{figure}

Our key result is contained in Fig.~\ref{fig3}, which displays the relative 
percentage deviation between the inhomogeneous Hubbard and the 
inhomogeneous Heisenberg models with $J'=J(\bar{U})$ for each of the four 
averages. The solid line represents the corresponding deviation obtained for
the impurity-free Hubbard and Heisenberg models, where $U$ and $J=4t^2/U$
are the same across the system. 

Figure~\ref{fig3}--(a) demonstrates clearly that if the background
$U$ is so small that the mapping is not quantitatively reliable even in 
the homogeneous system (this includes the values found in cuprates), then 
it is also not reliable in inhomogeneous systems, regardless of the choice
made for relating the bond defect to the impurity size. On the other hand, 
Fig.~\ref{fig3}--(b) makes a different statement: once the background $U$ is 
large enough to permit the basic Hubbard-to-Heisenberg mapping to function,
all four averages lead to values of $J'=4t^2/\bar{U}$ that are equal
(H) or different (A,G,Q) than obtained in the homogeneous system. 

This is unequivocal numerical evidence that the mapping of strongly interacting
fermions on weakly interacting spins survives in the presence of impurities
and defects. In order to probe the locality of the mapping we have
also performed numerical experiments with averages over more than two
neighboring sites, calculating $J'(\bar{U})$ from, {\em e.g.} a weighted 
average of the interactions at the sites connected by $U$ and their
nearest-neighbor sites. No improvement (and frequently even worse results)
with regard to the simple two-site averages was obtained, indicating that
the mapping is indeed local.

At first sight more surprising is that the alternative averaging procedures 
produce even smaller deviations for some values of $U$, Fig.~\ref{fig3}--(b). However, for still larger
values of $U$ and $U'$ (where the mapping should as a matter of principle
get better and better) all these alternative averages produce deviations 
that continue to grow and yield $D>0$, while the harmonic mean correctly 
approaches the limit $D=0$ as $U,U'\to \infty$. This shows that only the 
harmonic mean has a chance to correctly describe the fermion-to-spins mapping 
in inhomogeneous systems, while the lower deviations of the other averages 
for some parameter regimes only occur because the curves are monotonous so 
that there is always a range of values for which they are close to zero.

\begin{figure}[t!]
\includegraphics[height=5cm]{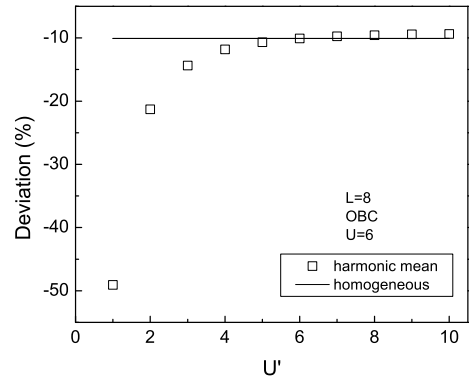}
\caption{Deviation of GS energies of the half-filled Hubbard model and the corresponding
Heisenberg model as a function of impurity strength, obtained
from the harmonic mean. At $U'=U=6$ the system
is homogeneous (except for the presence of the boundary) and the deviation
is the same obtained in Sec.~\ref{rate}.}
\label{fig4}
\end{figure}

We note that in Fig.~\ref{fig3} the ratio of $U$ and $U'$ was held fixed,
such as to guarantee that for all values of $U$, $U'$ was always substantially
different from $U$ ($U'=3U/2$). In Fig.~\ref{fig4} we present the complementary analysis 
in which the background $U$ is held fixed and the impurity interaction is 
varied from $U'<U$ to $U'>U$. For this comparison we consistently adopted
the harmonic mean. A first feature that jumps to the eye is that a single site
with $U'<U$ is enough to substantially deteriorate the mapping. By contrast,
a single site with $U'>U$ leads only to a slight reduction of the deviation
between GS energies, much less than the deterioration observed for $U'<U$. 

This behavior arises from the hopping terms. Both at $U' >U $ (more
repulsive impurity site) and at $U' <U$ (less repulsive site) the
on-site density at the impurity site slightly deviates from that at
the background sites, as long as $U$ and $U'$ are both finite. In the
latter case, however, hopping processes involving the impurity site
increase as $U'$ is reduced, and the Hubbard-to-Heisenberg mapping
becomes correspondingly worse, while in the former case hopping
continues to be strongly suppressed. The behavior displayed by Fig.
\ref{fig4} is thus consistent with what one would expect on the basis
of the derivation leading from Eq.(\ref{hubbard_inh}) to Eq.(\ref{inhomhe}).   

Independently of, but in 
agreement with, our previous analytical derivation we thus find that the harmonic mean
solves the problem exactly for a single impurity and for sufficiently large interactions,
while the other possible averages do not. However there is still one question concerning this result 
and it is addressed in next section: is the harmonic mean able to recover the mapping in more 
complex inhomogeneities? 

\subsection{More complex inhomogeneities}
\label{complexinhom}

In view of our initial discussion of naturally occurring or man-made 
inhomogeneity in strongly correlated systems, it becomes important to
extend our analysis to more complex inhomogeneities than boundaries or
single impurities. We here briefly describe our findings on three of
these: interfaces, superlattices and disordered systems.

\begin{figure}[t!]
\includegraphics[viewport= 50 50 300 800, angle=-90, width=10.5cm]{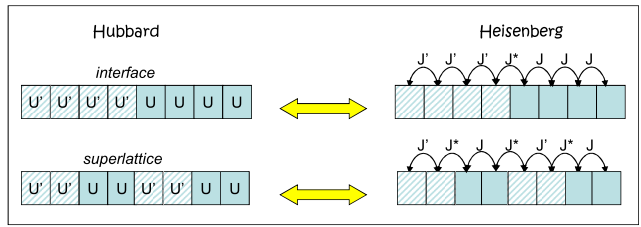}
\caption{Illustration of simple interface and superlattice structures in
the Hubbard model, and their counterparts in the Heisenberg model.}
\label{fig5}
\end{figure}

{\em Interfaces and superlattices} can be described in the Hubbard and 
Heisenberg models as shown schematically in Fig.~\ref{fig5}. While
the description of superlattices by means of periodic spatial variations
of $U$ is the standard choice,\cite{paiva} which we here also adopt,
it has been pointed out that a periodic modulation of local electric potentials
can bring about a much larger change in the system properties.\cite{malvezzi,malvezzi2,fc2008} 
Here, however, we are interested in the Hubbard-to-Heisenberg transition,
which is driven by the interaction and not by local potentials, and therefore
we follow the usual prescription to ignore possible local electric fields 
in the superlattice structure.

\begin{figure}[t!]
\includegraphics[height=6cm]{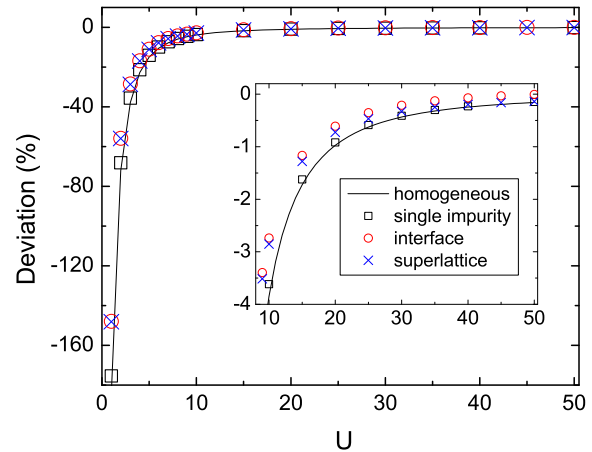}
\caption{Relative percentage deviation between the GS energies of the
Hubbard and the Heisenberg model for a superlattice structure, an interface,
a single-impurity system and a system that is spatially homogeneous (except
for the surface). In the superlattice and
the interface system the number of sites with $U$ and $U'$ is the same,
respectively, the only difference being in their geometric distribution.
System parameters: $L=8$, $U'=3U/2$, $J'=4t^2/U'$ and $J^*=4t^2/\bar{U}$,
where $\bar{U}$ is calculated from the harmonic mean of $U$ and $U'$.}
\label{fig6}
\end{figure}

We note that in the superlattice and the interface geometry we now have 
{\em three} different spin-spin interactions, one, $J$, being calculated from 
$U$, another, $J'$, from $U'$ and the last, $J^*$, from $U$ and $U'$, as 
indicated in Fig.~\ref{fig5}. Figure~\ref{fig6} shows that for large $U$ and $U'$ the GS energies of the
Heisenberg model, when calculated from the harmonic mean, become
identical to those of the corresponding Hubbard model, for all investigated
geometries. In this sense, the mapping continues to work and to be as local as
possible. Note that this would not be true for the arithmetic, geometric
and quadratic means, whose deviations increase for large $U$ and $U'$.

The fact that the Hubbard-Heisenberg deviation for the superlattice is larger 
than that for the interface can be understood on purely geometric grounds, 
as a consequence of the locality of
the mapping and the nature of the harmonic mean: By comparing the
distribution of $J$ values, Fig.~\ref{fig5}, we see that in going from the interface to the
superlattice the number of interactions $J$ and $J'$ is reduced by the
same amount, while that of interactions $J^*$ increases. Since $J' < J^* < J$,
a reduction of the number of sites with $J'$ worsens the mapping while a 
reduction of the number of sites with $J$ improves it. To see what the net 
effect is we must take into account the interaction $J^*$, which replaces $J$ 
and $J'$. This interaction is calculated from $\bar{U}$, the harmonic mean 
of $U$ and $U'$. The harmonic mean of any two positive numbers is less or 
equal their arithmetic mean, so that $\bar{U}$ is closer to $U$ than to $U'$ 
and $J^*$ is closer to $J$ than to $J'$. The substitution of an equal number 
of $J$ and $J'$ by $J^*$ thus has the effect of effectively increasing the 
number of `bad' sites, and therefore to deteriorate the quality of the 
mapping. This is what the data show: the deviation for the superlattice is
larger than that for the interfaces, if all other parameters are chosen the 
same.

This analysis shows that the harmonic-mean prescription
continues to be usable for these more complex geometries and that the
effect of the geometrical distribution of $U$ and $U'$ sites across the
system can be understood and analyzed essentially on a site-by-site 
basis. This is a direct consequence of the locality of the mapping.

{\em Disordered systems} can be modeled simply by considering a random
distribution of impurities instead of just one. While a complete analysis
of disordered systems requires statistical analysis of data resulting
from a large number of realizations of the disorder, an analysis of
a few representative cases is enough to conclude that the single-impurity
results are not changed, in their essence, when the impurity concentration
is increased.

Specifically, we find that if all impurities have $U'>U$ a higher concentration
of impurities reduces the deviation between the GS energies. Keeping the
concentration fixed and increasing $U'/U$ also reduces the deviation,
but to a much smaller degree. On the other hand, if all impurities have 
$U'<U$ the agreement is naturally worsened. However, in the $U'<U$ case the 
decisive factor is not so much the concentration of impurities but their 
strength, as measured by $U'/U$.
This inversion in the effect of concentration
and strength of the impurities can be understood on the basis of 
Fig.~\ref{fig4}, which shows that for a single impurity with $U'<U$ 
the deviation rapidly increases as $U'$ becomes more different from $U$,
while for $U'>U$ it decreases only very slowly and almost saturates as the
impurity sites effectively drop out of the system.

This discussion shows that it is possible to control the degree of the 
fermion-spin mapping by means of the introduction of a suitable concentration 
of impurities of suitable strength. This possibility may be useful in the 
design of nanoscale devices based on strongly correlated systems, whose
properties can be tailored from electron-like to spin-like by introducing
suitable disorder. We note in passing that this is a strong-interaction
effect, completely different from the itinerant-to-localized transition
resulting from disorder in Anderson localization.

\section{Conclusions}
\label{concl}

In the homogeneous situation, in which all sites are equivalent, the mapping is characterized by the 
behavior of the system as a function of the onsite interaction $U$.
Not unexpectedly, the Heisenberg model is found to be a good approximation to 
the t-J and the Hubbard model at $n=1$. Somewhat more unexpected is that
the Heisenberg model is rather a
bad approximation to the Hubbard model at $n=1$ even for values of $U$ that are
considered strongly correlated in solid-state applications. Only at $U$
near 20 has the deviation dropped to about one percent. The standard mapping
thus only becomes quantitatively reliable for values of $U$ that are
hard to reach in the solid state, but have already been demonstrated
in cold-atom systems.

In inhomogeneous situations, translational symmetry is broken.
An analytical calculation for the simple case of
a single impurity suggests that
the mapping can be preserved in terms of the harmonic mean. Moreover, in
terms of this mean the mapping is as local as possible, {\em i.e.,} the value 
of the Heisenberg $J$ between two sites is only determined from the value of 
the Hubbard $U$ at these two sites. Numerical calculations illustrate and
corroborate this finding. This is more than a mathematical, or numerical,
result: it means that the physics of the mapping, {\em i.e.}, the
gradual freezing out of the charge-degrees of freedom and the localization
arising from the concomitant suppression of double occupation, is essentially
the same regardless of the geometry and the presence or absence of
translational symmetry.

The harmonic-mean prescription can be used easily and reliably for a wide 
variety of spatial inhomogeneities. Once the basic mapping is understood, the
harmonic-mean prescription allows one to interpret and even to predict the 
behavior of much more complicated systems, without going through detailed
analytical or computationally expensive numerical calculations.

{\bf Acknowledgments} 
This work was supported by Brazilian funding agencies FAPESP and CNPq.
We thank Vivaldo Campo Jr. for providing us with his exact diagonalization
code for the Hubbard model, and Fabiano C. Souza and Valter L. L\'ibero 
for providing us with their exact diagonalization code for the Heisenberg model.

\end{document}